# Engaging the scientific community in high-quality biocuration: a report on the International Society for Biocuration workshop, ‘Maximizing community curation for the benefit of all’

Daniela Raciti[1], Susan L.M. Coort[2], Christian Grove[1], Jade Hotchkiss[3], Matt Jeffryes[4], Nancy T. Li[5], Zhiyong Lu[6], Bastien Molcrette[7,8], Sushma Naithani[9], Maria Victoria Nugnes[10], Jolene Ramsey[11], Rene Ranzinger[12], Leonore Reiser[13], Karen E. Ross[14], Garrett Stevens[15], Courtney Thaxton[16], Sabrina Toro[17], Valerie Wood[18], Karen Yook[1], Kimberly Van Auken[1]

**Affiliations**

1. Division of Biology and Biological Engineering, California Institute of Technology, 1200 E. California Boulevard, Pasadena, CA 91125, United States
2. Department of Translational Genomics, NUTRIM Institute of Nutrition and Translational Research in Metabolism, Maastricht University, Maastricht, the Netherlands
3. Division of Human Genetics, Department of Pathology, University of Cape Town, South Africa
4. Literature Services, EMBL-EBI, Wellcome Trust Genome Campus, Cambridge, United Kingdom
5. Ontario Institute for Cancer Research, Toronto, ON M5G 0A3, Canada
6. Division of Intramural Research (DIR), National Library of Medicine (NLM), National Institutes of Health (NIH), Bethesda, MD 20894, United States
7. GigaScience Press, BGI Hong Kong Tech Co Ltd., 26F A Kings Wing Plaza 2, 1 On Kwan Street, Shek Mun, Sha Tin, NT, Hong Kong SAR
8. Present address: Department of Chemistry, University of Basel, Building 1096, Mattenstrasse 22, CH-4058 Basel, Switzerland
9. Oregon State University, Corvallis, OR 97331, United States
10. Department of Biomedical Sciences, University of Padova, Padova 35131, Italy
11. Department of Biology, Center for Phage Technology, Texas A&M University, College Station, TX, United States
12. Complex Carbohydrate Research Center, The University of Georgia, 314 Riverbend Rd, Athens, GA 30602, United States
13. Phoenix Bioinformatics, 39899 Balentine Drive, Suite 200, Newark, CA 94560, United States
14. Georgetown University Medical Center, Washington, DC 20057, United States

15. Department of Bioengineering, University of California, Berkeley, Berkeley, CA, United States
16. Department of Genetics, University of North Carolina at Chapel Hill, Chapel Hill, NC, United States
17. The University of North Carolina at Chapel Hill, 120 Mason Farm Road, 5000 D Genetic Medicine Building CB#7264, Chapel Hill, NC 27599-7264, United States
18. Department of Biochemistry, University of Cambridge, Cambridge CB2 1GA, United Kingdom

## Abstract

Biological knowledgebases traditionally rely on expert, professional curation of the research literature to maintain up-to-date collections of data organized in machine-readable form. However, despite the increasing amount of curatable biomedical knowledge, support for knowledgebases is declining, leaving these resources no alternative but to explore additional ways of updating and maintaining content. One way in which knowledgebases have addressed this problem is by engaging researchers to help curate their published papers, a process generally known as 'community curation'. As helpful as community curation can be, though, it is not universally adopted and, for groups that do have it, there is a wide range of approaches.

To learn about existing community curation pipelines and explore possibilities for working towards a common approach, we organized a workshop, Maximizing Community Curation for the Benefit of All, at the 18th International Biocuration Conference, hosted by the Stowers Institute for Medical Research. Our aim was to examine the different strategies that groups use, share successes, failures, and ongoing challenges, and produce suggested deliverables for broader adoption of common best practices and tools for effective community curation.

Representatives from 18 different resources, ranging from model organism and specialty knowledgebases to journals and literature resources, presented their work. The result was a comprehensive assessment of the state-of-the-art for community curation and an in-depth discussion on how community curation can become standard practice for maintaining timely, high-quality biological resources that will continue to provide scientists with the essential information they need for their research.

# 1. Introduction

Biological knowledgebases are essential scholarly resources for the biomedical and agricultural research communities, providing readily accessible, expertly curated, structured data that drive novel discoveries, enable large-scale computational analyses, and support decision-making for clinical or other contexts [1,2]. Knowledgebases underpin a wide range of applications (e.g., functional genomics, systems biology, drug development, precision medicine, plant breeding, and agriculture). However, the rapid and continuous growth of the scientific literature, fueled by high-throughput technologies and collaborative, interdisciplinary research, has placed immense pressure on the traditional approach of manual, expert professional biocuration. With increasingly limited staff and decreasing resources, knowledgebases face mounting challenges in keeping pace with the accelerating volume and complexity of biological data [3-10].

As one strategy to address these challenges, community curation has gained traction as a complementary and scalable approach for knowledgebases to stay current [6,11-18]. Community curation enlists a distributed network of scientific contributors, beyond the core professional knowledgebase staff, to participate in the identification, extraction, annotation, and quality control of biological information from the scientific literature and other data sources. By engaging a broad pool of contributors—including researchers, clinicians, students, data scientists, and citizen scientists— community curation expands the scope and depth of curation efforts by leveraging collective scientific domain expertise while fostering a sense of shared responsibility for populating and maintaining valuable research resources.

Against this backdrop, we organized a community curation workshop, “Maximizing community curation for the benefit of all”, at the 18th International Society for Biocuration (ISB) Conference held at the Stowers Institute in Kansas City, Missouri, USA. The workshop convened a multidisciplinary group of knowledgebase stakeholders including biocurators, ontology developers, software engineers, and educators to discuss strategies for recruiting, training, and sustaining community contributors, common barriers to participation, and collaborative approaches and recommendations for building a more sustainable and impactful community curation ecosystem. This report highlights the insights and outcomes of the workshop and offers a roadmap for advancing community curation as an essential component of modern biological knowledgebase infrastructure.

# 2. Workshop structure and goals: drawing on the collective expertise of diverse biocuration resources

The 2025 ISB community curation workshop drew expert participants representing 18 biocuration resources including long-standing knowledgebases with established community curation pipelines as well as emerging platforms exploring the incorporation of community-based curation models (Table 1). The workshop was conceived as an opportunity for shared learning and strategic planning, recognizing the pressing need to scale biocuration through structured, accessible, and sustainable community engagement.

The workshop consisted of a series of short presentations in which each invited speaker highlighted their resource’s experiences with community curation, including their target community, curation tools, data types curated, challenges encountered, and incentives that successfully encourage participation. To guide the discussions, speakers were asked in advance to address these specific aspects, focusing not only on strategies that worked well but also on challenges so that participants could learn equally from successes and setbacks. Following the presentations, participants engaged in a collaborative discussion to identify insights, areas of convergence, and opportunities for improvement, revealing several unifying themes across the diverse resources represented and providing a framework for continued development of community curation approaches.

## 2.1 Participating biocuration resources

Biocuration resources are diverse and serve the research needs of a wide range of scientific communities. Therefore, when organizing this workshop, we included a broad spectrum of resources to represent the experiences and expertise of the greater biocuration community (Table 1). We aimed to maximize the opportunities for productive exchange of different ideas on how to scale biocuration through accessible and sustainable community engagement.

### Model Organism Knowledgebases

Some of the major biocuration resources that engage in community curation focus on cataloguing the genetics and genomics of commonly studied model organisms and related species. These model organism knowledgebases (traditionally known as Model Organism Databases (MODs)) typically have long-standing relationships with their user communities and are thus a natural fit for community curation approaches. Participating in this workshop were PomBase, the model organism knowledgebase for the fission yeast *Schizosaccharomyces pombe* [10], The Arabidopsis Information Resource (TAIR) [17], and WormBase, the knowledgebase for *C. elegans* and related nematodes [19], a founding member of The Alliance of Genome Resources [8], and originator of the ACKnowledge community curation platform [13].

### Multi-organism Knowledgebases

Other biological knowledgebases are dedicated to curating scientific knowledge across a wider variety of species, sometimes with a focus on a particular subject matter. Representatives from this group included the Universal Protein Resource (UniProt), a comprehensive database of protein sequence and function [20], DisProt, a database of intrinsically disordered proteins and regions [18], and GlyGen, a glycoinformatics knowledgebase with extensive glycoprotein and glycan data [21]. Plant-focused multi-species resources included Gramene [22], a comparative plant genomic resource, and the Plant Reactome pathway knowledgebase [23].

### Clinical/Disease Knowledgebases

Knowledgebases that focus on clinically relevant data also participated in the workshop. These included the Clinical Genome Resource (ClinGen), a resource of clinically relevant genetic knowledge [24], the Sickle Cell Disease Ontology (SCDO), a knowledgebase for researchers, patients and clinicians in the Sickle Cell Disease (SCD) field [25], and the Monarch Initiative [26], a collaborative project that integrates genetic, phenotypic, and disease data across species to support disease diagnosis, gene discovery, and translational research.

### Pathway Knowledgebases

Pathway knowledgebases are dedicated to the curation of biological pathways and processes. Among the resources represented at the workshop were Reactome, a peer-reviewed pathway knowledgebase in which human pathways are authored by domain experts [27], as well as Plant Reactome, which contains pathways for 139 species of green plant lineage [23], and WikiPathways, a community-driven platform that enables collaborative creation and curation of biological pathway models for 27 species [28].

### Undergraduate Academic Coursework

Engaging undergraduate students in biocuration is an additional approach to training and motivating the next generation of community curators. To learn more about these approaches, we heard from representatives of The Community Assessment of Community Annotation with Ontologies (CACAO), a crowdsourced undergraduate Gene Ontology (GO) community curation project [29], and WormBase, who has partnered with undergraduate institutions to curate *C. elegans* phenotypes [30]. Plant Reactome has also trained undergraduate students in gene and pathway biocuration [31].

### Genome Annotation

As costs for whole-genome sequencing have declined, the number of new genome sequences requiring annotation has increased. For many organisms, genome annotation is community-driven, and the availability of tools for performing the annotation is critical for timely analysis. To represent this type of community curation, we included a presentation on the Apollo collaborative genome annotation tool [32].

### Scientific Journals/Publishers

While most biocuration occurs post-publication, there are successful publishing models that integrate community curation into the pre-publication process. Examples represented in our workshop were GigaScience, where authors help to curate metadata for their datasets with guidance from professional curators at GigaDB (an open database managed by GigaScience) [33], and microPublication Biology and the Genetics and G3 journals published by the Genetics Society of America (GSA) that integrate entity recognition and validation either by the authors or by professional biocurators [34, 35], and the journals published by the American Society of Plant Biology (ASPB) (Plant Cell and Plant Physiology), who engage authors in improving gene annotation during the publication process [36].

### Literature Databases

Literature databases are key players in the biocuration ecosystem. The central bibliographic repositories for biomedical literature abstracts and full text (when available), PubMed, PubMed Central (PMC), Europe PMC [37], and the BioNLP research group at the National Library of Medicine, provide critical resources and state-of-the-art tools such as PubTator [38], LitSuggest [39], and LitVar [40] to help with the biocuration tasks of entity and concept recognition and document triage. We included representatives from these resources to understand how they could further help community curation.

## 2.2 Key discussion topics

The workshop was designed to explore six key aspects of community curation through structured presentations and discussion. These topics, described below, encompassed the full lifecycle of community curation initiatives, from identifying and recruiting contributors to ensuring long-term sustainability.

a) **Finding and engaging community curators:** Identifying and engaging community members to participate is one of the first steps in successful community curation initiatives. We were interested to learn how different resources recruit community members to participate and what has, or has not, been successful.
b) **Types of data community members curate:** The suitability of different data types for community contribution was also addressed. Understanding what types of data can be curated to existing high standards by the community is crucial to establishing and maintaining a viable community curation program.
c) **Quality assurance:** Integrating community-contributed data into knowledgebases requires suitable quality assurance. We wanted to learn what quality control mechanisms different resources employ and how they balance maintaining data integrity with reducing barriers to participation.
d) **Incentivization and recognition:** Attracting community curators is only the first step; retaining and keeping them engaged requires continuous outreach, thoughtful approaches to incentivization, and recognition of effort. We were interested to learn more about the range of successful recognition models.
e) **Machine learning and artificial intelligence:** Natural language processing pipelines, named entity recognition (NER) systems, document classification, and, more recently, large language models (LLMs) enhance the scalability of biocuration. We were interested in exploring how these approaches could help to shift the role of community contributors from de novo annotation to validation of machine-assisted pipelines.
f) **Sustainability:** Sustainability of community curation initiatives, and biocuration more generally, is a cross-cutting concern. We wished to learn how groups have, or plan to, enact stable community curation infrastructure, coordinate across resources, and adopt common best practices in community curation.

By including a wide range of participants and focusing discussion on several critical aspects of community curation pipelines, we hoped to generate a comprehensive assessment of the state of community curation for biomedical and plant-related resources, highlighting not only successes for broader adoption but potential areas of improvement and future collaboration. Our aim, overall, is to promote a maximally beneficial and sustainable role for community curators in the larger biocuration ecosystem.

# Section 3 - Key Discussions and Takeaways

**a) Finding and engaging community curators.**

Finding and engaging community members is a critical step in any successful community curation pipeline. While users can provide feedback to most resources online, for example via a 'Contact us' link, in our workshop we wished to explore more proactive and detailed strategies that groups use to engage community curators. Groups employ diverse tactics for identifying and engaging community curators, from volunteer expert panels to student competitions, each with distinct strengths and challenges (Table 2). Depending on the resource, contributors include authors of recent publications, subject matter experts, clinical researchers, bioinformaticians, graduate and undergraduate students, educators, or even citizen scientists. However, all workshop participants emphasized the importance of aligning engagement efforts with contributors' expertise and their motivations for participation. For example, model organism databases, such as PomBase, TAIR, and WormBase, have long-standing relationships with their user communities who recognise the importance of maintaining the knowledgebases they use on a daily basis and are therefore willing to curate their own publications. These resources commonly use direct email outreach to publication authors, with PomBase achieving response rates as high as 57% [10], and WormBase reporting that contacting all authors seems more effective than contacting only corresponding authors. Timing of contact matters significantly; uptake is highest immediately after publication when researchers are still actively engaged with the work, and especially when students and postdoctoral fellows are still in the relevant laboratory. At TAIR, however, authors were less responsive when asked to contribute directly after publication, but more submissions were received when community curation was part of the publishing pipeline (see below).

For other resources, community participation was a founding principle of their curation model, with domain experts contributing knowledge in close collaboration with professional biocurators. Reactome's approach emphasizes structured, reaction-level pathway representations, which require substantial domain expertise and familiarity with data models, which limits the extent to which contributions can be easily modularized and crowdsourced. More recently, Reactome has observed a shift from predominantly biocurator-initiated outreach to an increasing number of contributor-initiated contacts, suggesting that improved documentation and instructional resources (see below) can meaningfully lower the barrier to community participation. Plant Reactome engages researchers and community experts in pathway review using professional networking and outreach at conferences. GlyGen also relies on professional networking to solicit contributions via their website, while ClinGen recruits genetics professionals and clinicians through professional networks, conferences, and social media.

To expand and strengthen its curator base, WikiPathways provides a prominent 'Contribute' link on their website for community members to sign up as a WikiPathways Author, and actively engages specific research communities, such as the COVID-19 Disease Map project, LIPID MAPS, and rare disease networks [15], with online and in-person curation meetings.

SCDO engages subject matter experts with a vested interest in creating and maintaining valuable resources and mobilizes existing expert communities through collaborative working groups. For more specialized tasks, such as writing ontology term labels and definitions in layperson terms or reviewing auto-translations of these annotations into other languages, SCDO approaches relevant experts with the specialized skills and experience, such as SCD genetic counselors and bilingual technical experts.

The Monarch Initiative solicits, and is contacted by, experts to help refine specific areas of the ontologies. For example, Monarch has collaborated with experts via in-person and virtual workshops to improve prenatal phenotype representation of the human phenotype ontology (HPO) [41], and epilepsy representation in the Mondo disease ontology [42]. This approach is also employed by the GO Consortium to develop and refine domain-specific areas of GO, such as the extracellular matrix and multi-organism interactions, in concert with subject matter experts [43].

DisProt recruits through networking in biocomputing lab projects and conferences targeting subject matter experts, and encourages participation by providing training materials on curation best practices [44]. Additionally, in 2025, DisProt became a hybrid resource that serves both as a knowledgebase and as a deposition platform, accepting contributions from the broader scientific community and creating new opportunities to engage with users.

Academic courses provide an alternative recruiting model, with CACAO [29], WormBase [19], and Plant Reactome [23] successfully engaging undergraduate students through coursework, experiential learning opportunities, and/or paid internships, creating a renewable pool of contributors. In recruiting for CACAO, students are motivated to complete course work resulting in measurable and authentic contributions to the scientific research community on topic areas of their interest, rather than assignments only graded by instructors.

Some publishers have successfully integrated community curation into publication workflows. GigaScience [33]), microPublication Biology [34] and journals published by the GSA [35] and the ASPB [36] require, or have required, author-assisted data curation as part of the publication process.

Other resources, such as Europe PMC and the National Library of Medicine (NLM) do not have a traditional community curation program but rather provide essential infrastructure and AI tools such as LitSuggest [39] and BioC [45] to support curation across the ecosystem.

Despite these varied approaches, a common challenge emerged across nearly all projects: the participation barrier remains significant. Even though the majority of resources make contribution opportunities highly visible through prominent website buttons and integration into existing tools and workflows, many potential contributors are still unaware of curation opportunities and benefits or find the barrier to participation too high, even when user-friendly tools and training are provided.

### b) Types of data community members curate and curation tools.

An equally important consideration is determining which types of data are best suited for community curation (Table 3). This decision depends on both the needs of the research community, such as which data are currently of highest priority for discovery, and practical factors including data complexity, task clarity, available training, and infrastructure support.

Across resources, community members contribute to diverse data types: from gene and protein function annotations (ACKnowledge, CACAO, DisProt, PomBase, TAIR, UniProt, Plant Reactome, WikiPathways), to clinically relevant variants (ClinGen), metabolic and signaling pathways (Plant Reactome, Reactome, WikiPathways), phenotypes (Monarch Initiative, PomBase, SCDO, WormBase), and genome annotation (Apollo, Gramene, TAIR), as well as ontologies (Monarch Initiative, SCDO). Other resources focus on topic–specific curation, such as glycobiology (GlyGen), while publishers focus on varied datasets from recently accepted manuscripts (GigaScience, microPublication Biology).

One common approach that doesn't require intensive training is to ask authors to 'triage' or 'flag' the experimentally studied entities (e.g., genes) and data types (e.g., mutant phenotypes) reported in their publications. Alternatively, some resources further encourage authors to provide comprehensive curation of all data types in their publications to a professional standard. Other resources engage the community in curating pathway models, genome annotations, disease-associated genomic variants, and domain-specific ontologies.

Some resources, like Europe PMC, PubTator, and LitVar, allow users to provide feedback on their machine learning approaches to identify and link biological entities, such as genes, proteins, diseases, chemicals, and organisms, to publications.

Inherent in curation of different data types is varying levels of biological and, thus, curation complexity. Participants noted a recurring tension between data complexity and curation tool usability. Expert, professional biocurators are comfortable navigating complex curation tools, but community contributors may need or prefer a simpler user interface, at least initially.

In some cases, use of familiar tools is sufficient for community curation tasks. With the SCDO's undertaking to translate its term labels and definitions into layperson terms and into other languages relevant to the SCD field, it makes use of tools, such as Google spreadsheets. This approach supports a more centralized biocuration workflow, offline editing, ready tracking of provenance, and facilitates frequent public releases to the SCDO.

More customized interfaces, such as PomBase's wizard-style guided curation tool (Canto) [46], the ACKnowledge AI-assisted curation tool [13], and the microPublication Biology entity recognition tool, have successfully lowered barriers to community participation. These platforms offer contributors user-friendly interfaces that are thorough but don't unnecessarily expose community curators to underlying database models, which in some cases, may be quite complex. Such user-friendly tools, however, are still underpinned by robust, standards-compliant pipelines that ensure high-quality, interoperable data. Continuous improvement to such systems to accommodate different user roles and levels of engagement was identified as an important area for ongoing development.

Plant Reactome has successfully incorporated both types of tools in a tiered approach to community curation. Initially, students and community curators use Google spreadsheets for gene function and gene and pathway summaries, but as they gain more experience graduate to using tools such as WikiPathways [28] and then ultimately to the Reactome Curator Tool [31].

As with professional biocuration, engaging the community in curation of different data types of varying complexity is important for supporting the efforts of a knowledgebase or ontology, and having access to efficient, user-friendly tools is clearly an important factor in achieving this aim.

**c) Quality assurance.**

Ensuring the reliability of community-contributed data depends on effective quality assurance strategies. High-quality community curation not only ensures the reliable integration of data into knowledgebases but also promotes FAIR principles and responsible practices in the publication of experimental results, ultimately enhancing reproducibility, accessibility, and trust in shared scientific resources. Striking the right balance between rigorous quality control and low barriers to participation is critical, particularly given the wide range of contributor expertise and the competing demands on researchers' time. While professional biocurators remain central to quality assurance for all resources, providing expert review, ontology editing, and oversight of submissions, to better understand how different resources further address quality assurance, we explored the various mechanisms used to ensure data accuracy (Table 4).

Participants described a spectrum of strategies ranging from training prior to curation, constraints on community curation interfaces, and consensus and expert review. Reactome has invested in lightweight onboarding materials aimed at prospective community contributors (https://reactome.org/community). These include instructional YouTube videos that walk contributors through the pathway authoring and review process. In addition, many tools provide FAQs and in-line 'tool tips' to guide contributors on data submission. DisProt provides training materials in multiple languages accessible through TeSS, the ELIXIR Training eSupport System [47], holds periodic community meetings, and offers specialized training sessions with experts. Additionally, all deposited data are reviewed by expert biocurators prior to integration, and existing annotations are periodically reassessed to ensure accuracy and currency.

Most, if not all, interfaces, such as those implemented by ACKnowledge, CACAO, GlyGen, PomBase, TAIR, microPublication Biology, and WormBase constrain contributor input with features such as entity autocomplete, which ensures correct use of nomenclature and database identifiers, and reduces data entry errors. Reactome and Plant Reactome provide a structured authorship template that guides experts in organizing pathway content in a curator-compatible format, reducing the cognitive and technical overhead associated with contributing high-quality annotations. Genome annotation using Apollo includes gene structural validation and version tracking with rollback capabilities.

Community contributions to ontologies, universally used but typically requiring more advanced biocuration knowledge and tools, involve collaborations with experienced ontology editors who help to create, define, and organize new terms, and use existing quality-control pipelines to ensure that logical constraints and appropriate metadata standards are met [48].

Some resources, such as CACAO, TAIR, Reactome, Plant Reactome and ClinGen, employ peer review and expert panel verification, ensuring that curated data meets established standards and is supported by consensus among domain experts. The Monarch Initiative and SCDO have also done this to a degree, with certain term annotations and relations between terms reviewed by expert volunteers from relevant working groups or domain experts. In addition, SCDO, when possible, has more than one community curator review annotation translations.

WikiPathways uses a distributed community curation model in which all edits are attributed to registered authors and independently reviewed by at least one other curator. To ensure consistency in the curation practices, WikiPathways uses a standardized curation protocol consisting of interactive tasks such as reviewing recent edits, identifying redundancy, and resolving problematic content. This approach ensures quality control while allowing the resource to scale with new biological discoveries and diverse sources of pathway knowledge.

Overall, quality assurance is an essential part of community curation, upholding scientific rigor and confidence in the submitted data, therefore all groups spend significant effort in ensuring this goal is met.

**d) Incentivization and recognition.**

While the benefits of community curation to biomedical resources may seem readily apparent , all groups recognize that community curation needs to be adequately incentivized and recognized to encourage and promote participation. We were interested to learn more about the range of successful approaches, including authorship credit, contribution tracking and display on websites, gamification, and integration with coursework for academic credit. A synopsis is presented in Table 5.

As most researchers are regular users of knowledgebases, they have a vested interest in keeping the content accurate and up to date. Having the opportunity to facilitate this through community contributions thus provides a strong incentive to participate. Several groups, such as PomBase and WormBase, expedite the incorporation of community-curated references so that the data from those submissions become the highest priority for further curation, quality control checks, and subsequent inclusion into the resource.

An additional incentive is the opportunity to engage more directly with knowledgebase curators to highlight new types of data and to promote dialogues that improve data representation (e.g., by clarifying ontology terms or drawing attention to other data in need of curation). Further, by participating, researchers become more proficient in knowledge curation, a professional and transferable skill, by exposure to key concepts in data science literacy, such as awareness of data formatting, metadata, and curation standards.

Increasing publication visibility is another key motivation for community curation. A common and easily implemented form of recognition is to acknowledge community contributions on relevant website pages. PomBase highlights exceptional curated publications through a front-page spotlight and archive, boosting their prominence and publication page accesses, and in a recent PomBase community survey, 88.5% of participants reported that curation increased the visibility of their work. WormBase also credits community-contributed data on individual reference pages.

One common way groups publicly recognize individual community contributors is by attributing annotations directly to them. The SCDO has pages on their website recognising people involved in the inaugural working groups that contributed to the initial development of the SCDO, and if knowledgebases have web pages for community members, such as WormBase, their contributions can additionally be listed there. Further, resources such as APICURON (see below), DisProt, the Monarch Initiative, Reactome, WikiPathways and UniProt, associate community contributions with an ORCID (Open Researcher and Contributor ID) providing formal recognition that can be included in academic profiles and Curricula Vitae (CVs).

Including community contributors as authors on relevant publications or other scholarly research outputs are additional strong incentives. In Reactome and Plant Reactome, formal scholarly credit is achieved by assigning a DOI to each pathway and providing fine-grained ORCID attribution for contributors acting as pathway authors or reviewers [49]. In addition, ClinGen, DisProt, Plant Reactome, SCDO, and Monarch Initiative recognize community contributors as authors on relevant resource publications.

For community curation that is part of the publication process, such as that employed by Gigascience, microPublication Biology, and the Genetics and G3 journals published by the GSA, incentive is directly tied to publishing results in a timely manner. Participation in these processes is typically mandatory for publication, and provides authors with an opportunity to ensure that their data is free from error prior to publication. In some cases, it also helps connect new data with the larger research ecosystem by, for example, online linking of studied entities such as genes and variations, to their respective knowledgebase pages.

For some resources, incentivization is provided by awarding educational credits for contributions. ClinGen, for example, provides certificates of training that may be included on CVs, resumes, and LinkedIn profiles, and contributions can be used to fulfill 30-100 credit hours of continuing medical education per year. Similarly, undergraduate curation efforts are typically tied to coursework for which students receive credits towards their degree and experiential learning. Examples of this approach include CACAO, curation of *C. elegans* phenotype data for WormBase, and Plant Reactome, where students may earn research credits, write honors theses, be paid with grant funds, and/or earn co-authorship on presentations and papers.

Gamification also incentivizes and rewards community curators for participation. DisProt has led the way in this area by developing the APICURON system that awards badges and medals for participation [50]. APICURON is a readily adapted, community-wide platform for tracking contributions of professional and community curators across multiple databases. Badges are awarded for specific curation milestones while medals acknowledge top-performing curators and long-term achievements. CACAO runs online intercollegiate annotation competitions for student teams with public leaderboards that incentivize accuracy and consistent engagement in the annotation process.

More tangible incentives have also proven to be effective. For example, top-performing CACAO students are awarded chocolate bars, TAIR participants received 20th anniversary pins, and at the International *C. elegans* Meeting, top contributors to WormBase via the ACKnowledge form were awarded “worm trophies”, an approach whose success was very popular.

In addition, curated data may be shared across resources, extending its reach. This is illustrated by Gene Ontology annotations that appear not only on the GO website, but on model organism knowledgebases, UniProt protein records, and NCBI gene pages [43]. This broader reach helps to serve as evidence of data dissemination which can be included in the data management sections of grant proposals. In the 2025 PomBase community survey, 147/618 respondents reported using community curation contributions in this way, and a further 247 intend to do so.

In sum, biological resources that promote community curation recognize that it is imperative to offer valuable incentives and multiple forms of recognition, both formal and informal, to acknowledge the contributions of their community and thank them for the service they provide.

**e) Use of automation, machine learning, and artificial intelligence.**

As a number of biocuration resources already incorporate automation and state-of-the-art machine learning and artificial intelligence (ML/AI) into their professional biocuration pipelines, it is natural to extend these tools and methods to community curation. ML/AI applications for community curation can assist in named entity recognition, document triage/classification, fact extraction, and ontology development, as summarized in Table 6.

Platforms such as ACKnowledge and microPublication Biology use entity recognition to prepopulate specific fields on community curation forms. ACKnowledge also employs machine learning–based methods to predict data types, such as gene expression and protein interactions, found in the paper. UniProt is experimenting with using LLMs and other entity recognition and relation extraction tools to generate draft annotations that can be sent to authors for verification. In both cases, authors validate the prepopulated data to expedite the curation process and their feedback is used to improve machine learning methods. Prepopulating forms also reduces the burden of participation and helps authors understand what information is critical for curation.

DisProt recently developed, in collaboration with the Swiss Institute of Bioinformatics (SIB) Text Mining, DisTriage, an AI-based literature triage tool that automatically ranks articles by relevance for curation and makes them available to DisProt community curators through weekly reports [18]. Also ClinGen, Reactome, and Plant Reactome use AI to identify papers suitable for curation. Reactome has also developed an LLM-assisted curation workflow that augments professional biocuration by prioritizing relevant genes, literature, and functional annotations, with the longer-term aim of informing scalable community-facing curation workflows [51]).

WikiPathways also employs AI/ML to assist with pathway curation, including pathway figure or diagram processing, to suggest identifiers (genes and proteins) and provide recommendations for pathway layout and representation.

The Monarch Initiative has developed several AI tools to support community contributions to ontology development and curation, such as OntoGPT [52] supporting extraction of structured information from text and grounding it in existing ontologies, and CurateGPT [53] which links generative AI to trusted knowledge bases and literature, allowing curators to search and integrate information with traceable supporting evidence.

The SCDO's translation process uses language translation software to produce auto-generated term labels and definitions, which are then reviewed by community curators. They are currently working on improving their workflow by automating the inclusion of existing translations from other bio-ontologies (currently only the Human Phenotype Ontology) who make their translations available in specially formatted (babelon) files [54]. This will speed up the review process for community curators, as they can reuse existing translations used by other reliable bio-ontologies. The SCDO will then reciprocate, making their own translation files available in babelon format.

LitSuggest is a web-based machine-learning system for literature recommendation and triage in PubMed that uses ensemble text-mining classifiers trained on curated PubMed articles and continuously improves through curator feedback [39]. In the EnzChemRED study [55] LitSuggest was used to identify enzyme-function papers for UniProtKB/Swiss-Prot [20] and Rhea curation workflows [56] with high precision and recall. LitSuggest is also used by resources such as the GWAS Catalogue to support scalable biomedical literature curation [57].

Other resources, such as PubTator and Europe PMC, use text mining and machine learning to identify entities (such as genes, chemicals, and organisms), concepts (e.g. Gene Ontology terms) and relations (e.g. associate, positively correlate), all of which could be incorporated into other community curation pipelines, with the prospect of feedback on their methods forming a mutually beneficial relationship [58]. Indeed, PubTator has been integrated into the production curation workflows across multiple databases, such as UniProt [59] and Comparative Toxicogenomics Databases (CTD) [60], for enhanced efficiency.

For genome annotations, Apollo can display predicted gene models generated by machine learning algorithms, such as those produced by Tiberius [61] or Helixer [62] Curators can then use these gene models either as a starting point for curating the genome annotations of a newly sequenced species or to expand and refine existing species' gene annotation sets.

Automation, machine learning, and AI applications provide opportunities for more efficient biocuration and can help guide community curators to focus on the entities and concepts most important for inclusion in knowledgebases and other resources. AI tools are developing rapidly and we fully anticipate that more resources will adopt AI into their professional and community curation workflows as a matter of standard practice.

**f) Sustainability.**

While community curation initiatives provide much-needed and appreciated contributions, they still require software engineers and professional biocurators to design and implement curation tools and provide training and quality assurance. Thus, in the current funding landscape, sustainability of high quality community curation pipelines remains a concern. Strategies adopted by participating groups are summarized in Table 7.

Diversifying and increasing funding for biocuration resources would help sustain the professional positions that support community curation. Embedding support for community curation within research grant proposals, student training grants, or open science represents a promising avenue for securing such funding. Institutional funding within higher education aimed at training students has provided resources for training materials and software and developer support for CACAO community curation infrastructure [29]. Collectively, these approaches have the potential to ensure that community-assisted curation becomes a routine part of scientific training.

Another approach is incorporating biocuration into professional continuing education, a forward-thinking initiative that has already been successful for some groups, such as ClinGen. As continuing education is a requirement for some professions, integrating community curation into this training affords a continuous source of curators who can contribute well past their initial training.

Additional sustainable avenues include partnerships with academic journals or professional societies to support mandatory pre-publication curation. The community curation pipelines employed by GigaScience, microPublication Biology, and the GSA are all successful examples of this approach, and expanding it to more journals would be a path towards increased sustainability. Although pre-publication curation incurs a cost in production time to the journal as well as changes in their workflow, the incentive is to avoid extra exchanges between reviewers, editors, and authors to resolve ambiguous, incorrect, or missing information (data, metadata, software) that can delay the publication process or result in post-publication corrections.

Centralizing and standardizing community curation tools and guidelines is another option for promoting sustainability as it reduces potentially redundant efforts. Developing and maintaining shared infrastructure, such as curation platforms with built-in quality control, databases, and contributor recognition mechanisms, can leverage the collective creativity and expertise of different biocuration resources. Sharing training materials and best practices guidelines would help with global adoption of consistent standards and promote community awareness of FAIR principles of data collection and dissemination. These approaches could also lead to a more consistent user experience if common tools and guidelines were re-used by multiple resources. A centralized community curation hub would also make it easier for new resources to adopt existing community curation pipelines. The International Society for Biocuration (ISB), who hosted this workshop, could be central to coordinating and/or promoting a common community curation system, with successful collaborative biocuration efforts, such as the Alliance of Genome Resources, the Gene Ontology, and the Monarch Initiative's OBO Academy serving as models for this approach.

Ultimately, ensuring the long-term sustainability of community curation will require a combination of these approaches rather than reliance on any single solution. Diversified funding streams, integration with training and continuing education, journal and society partnerships, and centralized infrastructure are complimentary approaches that could maintain the professional expertise and technical support that are needed for high-quality community curation. By building on existing models and leveraging the reach of the ISB, the biocuration community could make community curation a lasting, integral component of the broader scientific enterprise, ensuring that knowledge remains accessible, accurate, and useful for researchers.

# Section 4 - Actionable outcomes and concluding remarks

Biological knowledge resources provide accurate, high-quality, evidence-based assertions to support the research needs of the biomedical community. Professional biocuration underpins these resources, but as the biocuration landscape evolves, specifically in response to funding constraints and technological advances, the broad need for successful community curation pipelines becomes increasingly apparent. The goals of the ISB community curation workshop were: 1) to share information about existing community curation pipelines and 2) to outline concrete steps that biocuration resources can collectively take to create an efficient and sustainable community curation ecosystem.

Many workshop participants noted that the traditional funding streams that have long supported professional biocuration are declining. As the volume and complexity of biological data continue to expand, these funding pressures have created an increasing gap between published data and what professional biocuration teams can realistically curate. In this environment, community curation is not merely a complementary activity but an essential strategy for sustaining biological knowledgebases. Without scalable models for community involvement, many critical resources will struggle to maintain adequate coverage as research output accelerates.

The need to triage priority areas for biocuration in low funding environments motivated an emphasis on the value of shared infrastructure where software development by one resource could be readily adopted by others. For example, centralized dashboards or analytics tools that track community curation efforts across resources could help to reduce redundant efforts, particularly for multi-species papers. Further, shared tools to help identify un- or under-annotated areas, highlight potentially stale knowledgebase entries in need of review or update, and suggest curation priorities, could collectively target community efforts to areas in most need of curation. In addition, integrating community curation tools with literature repositories such as PubMed or Europe PMC could improve visibility and opportunity for participation by reaching researchers on platforms used daily for finding and exploring the published literature.

Although making community curation opportunities widely available with such links could help, proactive and continued engagement by professional biocurators still seems a key component of sustained participation. At WormBase, for example, a link on the homepage to contribute gene expression data was used only 40 times in 15 years, while authors contributed 446 gene expression submissions, an ~ten-fold increase, via the ACKnowledge pipeline that actively emails authors. This suggests that passive links to community curation forms on knowledgebase websites may be underutilized compared to having managed infrastructure that regularly contacts the community for participation.

Routine, sustained community contact and training by professional biocurators could also help to sustain participation by creating a scientific ecosystem in which community curation is an inherent part of the data lifecycle. This is supported by the PomBase experience in which trained community members who start to participate from their earliest publications generally continue. In addition, by involving undergraduates in curation efforts, the trained scientific workforce is more likely to use knowledgebases and contribute to biocuration efforts across their career. Further, investments in training community curators could help keep scientists engaged even if they transition to other projects or areas; the skills acquired for curation in one area could be transferred to another. Community curation also does not require researchers to still be active in the lab. Opportunities to contribute once trained scientists retire can provide a way for experts to stay engaged and contribute their valuable knowledge to resources.

Although the workshop discussed the use of machine learning and AI pipelines as they existed at the time, this area is developing rapidly and thus, the landscape and opportunities for using AI in community curation has radically changed since then. Rather than being an optional component of biocuration, AI is essential for efficient workflows and shifts the role of community curators from creating de novo annotations to reviewing AI-suggested ones. For example, verifying AI-extracted experimentally relevant entities, data types, and findings in a publication via pre-populated forms builds on existing approaches and provides a familiar entry point for AI-assisted community curation. More comprehensive approaches could use AI to generate summaries of biological knowledge, such as gene or pathway descriptions, that community curators could help to evaluate and refine. Professional biocurators would then shepherd high quality, community-vetted, AI-generated content into knowledgebases. Effective integration of AI will require transparent pipelines to ensure knowledge provenance, clear quality assurance frameworks, and facile mechanisms for human validation and feedback.

In summary, while community curation frameworks vary widely in their implementation and scope, several common themes emerged from the workshop discussions. Shared, centralized infrastructure to support a uniform curation experience, broader coordination amongst different resources to track data submissions, for example from multi-species papers, collective approaches to target high-value data in need of curation, and broader adoption of reliable AI pipelines to help scale curation are all areas in which the biocuration community could work together, with the support of the ISB, to address these challenges. The ISB already provides information on community curation opportunities on the Curate Now page (https://www.biocuration.org/curate-now/), however this page is mainly accessed by biocurators and is likely not well known to researchers. To help bridge this gap, the ISB could consider fostering closer relationships with professional societies, such as the GSA or the Society for Neuroscience, to begin to systematically train scientists in biocuration and knowledge management. Such an investment would help not only to sustain knowledgebases but to enhance scientists' professional development. As many professional societies sponsor scientific meetings, a presence at these meetings with promotional materials and hands-on workshops would raise awareness of opportunities for community curation.

Community curation, however, cannot exist without the stewardship of professional biocurators to engage the broader scientific community in the curation process and to maintain the high standards of curated data. We hope this workshop launches a productive collaboration among knowledge resources to work together to make high-quality community curation an integral part of the scientific data life cycle.

# 5. Author contributions

Daniela Raciti (Conceptualization , Writing – original draft , Writing – review & editing), Susan L.M. Coort (Writing – review & editing), Christian Grove (Writing – review & editing), Jade Hotchkiss (Writing – review & editing), Matt Jeffryes (Writing – review & editing), Nancy T. Li (Writing – review & editing), Zhiyong Lu (Writing – review & editing), Bastien Molcrette (Writing – review & editing), Sushma Naithani (Writing – review & editing), Maria Victoria Nugnes (Writing – review & editing), Jolene Ramsey (Writing – review & editing), Rene Ranzinger (Workshop presenter), Leonore Reiser (Writing – review & editing), Karen E. Ross (Writing – review & editing), Garrett Stevens (Writing – review & editing), Courtney Thaxton (Writing – review & editing), Sabrina Toro (Writing – review & editing), Valerie Wood (Writing – review & editing), Karen Yook (Writing – review & editing), Kimberly Van Auken (Conceptualization, Writing – original draft , Writing – review & editing ).

# 6. Group contributors list

The following individuals contributed to this work through their participation in the indicated projects and resources.

**ACKnowledge:** Valerio Arnaboldi (ORCID 0000-0002-2563-5374), Daniela Raciti (ORCID:0000-0002-4945-5837), Paul W. Sternberg (ORCID 0000-0002-7699-0173), and Kimberly Van Auken (ORCID 0000-0002-1706-4196). **Apollo:** Garrett Stevens (ORCID 0000-0002-9781-5323). **CACAO:** Deborah A. Siegele (ORCID 0000-0001-8935-0696), Jolene Ramsey (ORCID 0000-0002-3774-5896), and Curtis Ross (ORCID 0000-0003-3035-6940). **ClinGen:** Courtney Thaxton (0000-0002-6733-369X) and Pepper St. Clair (0009-0007-1636-9645). **DisProt:** Maria Victoria Nugnes (ORCID 0000-0001-8399-7907), Maria Cristina Aspromonte (ORCID 0000-0002-4937-6952), Kamel Eddine Adel Bouhraoua (ORCID 0000-0001-9531-6339), and Silvio C.E. Tosatto (ORCID 0000-0003-4525-7793). **Europe PMC:** Matt Jeffryes (ORCID 0000-0001-9868-6271), Melissa Harrison (ORCID:0000-0003-3523-4408). **GigaScience:** Bastien Molcrette (ORCID 0000-0002-5995-5376), Chris Hunter (ORCID: 0000-0002-1335-0881), Mary Ann Tuli (ORCID: 0000-0002-4667-9528), Yannan Fan (ORCID: 0000-0003-3308-6878), Chris Armit (ORCID: 0000-0002-9952-8141), and Scott C. Edmunds (ORCID: 0000-0001-6444-1436). **GlyGen:** Rene Ranzinger (ORCID 0000-0003-3147-448X). **microPublication Biology:** Daniela Raciti (ORCID:0000-0002-4945-5837), Karen Yook (ORCID 0000-0002-4457-6787), and Paul Sternberg (ORCID 0000-0002-7699-0173). **Monarch Initiative:** Sabrina Toro (ORCID 0000-

0002-4142-7153). **NLM:** Zhiyong Lu (ORCID 0000-0001-9998-916X). **Plant Reactome and Gramene:** Sushma Naithani (ORCID 0000-0001-7819-4552). **PomBase:** Valerie Wood (ORCID 0000-0001-6330-7526). **Reactome:** Nancy T. Li (ORCID 0000-0002-2663-5245). **Sickle Cell Disease Ontology (SCDO):** Jade Hotchkiss (ORCID 0000-0002-2193-0704), Victoria Nembaware (ORCID 0000-0001-7966-4042), Ambroise Wonkam (ORCID 0000-0003-1420-9051, and Nicola Mulder (ORCID 0000-0003-4905-0941). **TAIR:** Leonore Reiser (ORCID 0000-0003-0073-0858). **UniProt:** Karen E. Ross (ORCID 0000-0003-4633-6055). **WikiPathways:** Susan L.M. Coort (ORCID 0000-0003-1224-9690), Egon L. Willighagen (ORCID 0000-0001-7542-0286), Martina Kutmon (ORCID 0000-0002-7699-8191), Alexander Pico (ORCID 0000-0001-5706-2163), Kristina Hanspers (ORCID 0000-0001-5410-599X), Friederike Ehrhart (ORCID 0000-0002-7770-620X). **WormBase:** Christian Grove (ORCID 0000-0001-9076-6015) and Paul Sternberg (ORCID 0000-0002-7699-0173).

# 7. Funding

Institutional support provided to the participating databases and knowledgebases is as follows:

**ACKnowledge:** This work was supported by the U.S. Department of Health and Human Services, National Institutes of Health, National Library of Medicine [grant number R01 OD023041].

**Apollo:** This work was supported by the U.S. National Science Foundation [grant number 2031120]; the UK Biotechnology and Biological Sciences Research Council [grant number BB/T016299/1]; and the U.S. Department of Health and Human Services, National Institutes of Health, National Institute of General Medical Sciences [grant number R01 GM080203].

**The Alliance of Genome Resources:** This work was supported in part by the U.S. Department of Health and Human Services, National Institutes of Health, National Heart, Lung, and Blood Institute [grant number U24 HG010859]; and the U.S. Department of Health and Human Services, National Institutes of Health, National Human Genome Research Institute [grant number U24 HG010859].

**CACAO:** This work was supported by Texas A&M University.

**ClinGen:** This work was supported by the U.S. Department of Health and Human Services, National Institutes of Health, National Human Genome Research Institute [grant numbers U24 HG009650, U24 HG006834, and U24 HG009649]; U.S. Department of Health and Human Services, National Institutes of Health, National Cancer Institute [grant number U24 HG009649]; and affiliate grants by the U.S. Department of Health and Human Services, National Institutes of Health, National Human Genome Research Institute [grant numbers U24 HG013077 and U24 HG010615]; and the U.S. Department of Health and Human Services, National Institutes of Health, National Cancer Institute [grant number U24 CA275783].

**DisProt:** This work was supported by the European Union Horizon Twinning, project IDP2Biomed [grant number 101160233]; and the European Union Horizon Marie Skłodowska-Curie Actions, project IDPfun2 [grant number 101182949].

**Europe PMC:** This work was supported by 36 funders of life science research (https://europepmc.org/Funders/) under the Wellcome Trust [grant number 326323] and European Union, European Research Council [grant number 10.3030/101034194]; by the European Molecular Biology Laboratory-European Bioinformatics Institute to MH [Europe PMC 2026-2031]; M Jeffryes was supported by the ARISE project, which has received funding from the European Union Horizon Marie Skłodowska-Curie Actions [grant number 945405]

**GlyGen:** This work was supported by the U.S. Department of Health and Human Services, National Institutes of Health, National Institute of General Medical Sciences [grant number 1R24 GM146616-01]; and the U.S. Department of Health and Human Services, National Institutes of Health Office of Strategic Coordination – The Common Fund [grant number 1OT2OD032092].

**LitSuggest, LitVar, and Pubtator:** This work was partly supported by the U.S. Department of Health and Human Services, National Institutes of Health, National Library of Medicine, Intramural Research Program.

**The Monarch Initiative:** This work was supported by the U.S. Department of Health and Human Services, National Institutes of Health, Office of the Director [grant number 5R24 OD011883]; and the U.S. Department of Health and Human Services, National Institutes of Health, National Human Genome Research Institute [grant number 5RM1 HG010860].

**Plant Reactome:** This work was supported by Oregon State University; the U.S. National Science Foundation [grant number 2029854]; and the U.S. Department of Agriculture, Agricultural Research Service.

**PomBase:** This work was supported by the Wellcome Trust [grant number 218236/Z/19/Z].

**Reactome:** This work was supported by the U.S. Department of Health and Human Services, National Institutes of Health, National Human Genome Research Institute [grant numbers U24 HG012198 and U24 HG012198-02S1 (Core Trust Seal development), and U24 HG011851]; the European Molecular Biology Laboratory, European Bioinformatics Institute; Open Targets [grant number OTAR-006, 2015-2024]; and York University.

**Sickle Cell Disease Ontology:** This work was supported by the U.S. Department of Health and Human Services, National Institutes of Health, National Heart, Lung, and Blood Institute [grant number U24 HL135600].

**UniProt:** This work was supported by the U.S. Department of Health and Human Services, National Institutes of Health, National Human Genome Research Institute [grant number U24 HG007822].

**WikiPathways:** This work was supported by ELIXIR Europe; the European Food Safety Authority [grant number GP/EFSA/ED/2022/01]; Maastricht University; the Netherlands Research Council [grant number VHP4Safety NWA-ORC 1292.19.272]; and the U.S. Department of Health and Human Services, National Institutes of Health, National Institute of General Medical Sciences [grant numbers GM103504 and GM100039].

**WormBase:** This work was supported by the U.S. Department of Health and Human Services, National Institutes of Health, National Human Genome Research Institute [grant number U24 HG002223]; the UK Medical Research Council [grant number MR/L001020/1]; and the UK Biotechnology and Biological Sciences Research Council.

# 8. Acknowledgements

The workshop organizers, Daniela Raciti and Kimberly Van Auken, are deeply thankful to the invited speakers, panelists, and all participants whose insights, shared experiences, and active engagement shaped the substance and direction of this report. Their collective expertise and openness fostered a productive and collaborative environment for discussing the future of community curation. We further acknowledge the International Society for Biocuration whose continued commitment to advancing biocuration made this workshop possible.

# 9. References

1. Wood V, Sternberg PW, Lipshitz HD. Making biological knowledge useful for humans and machines. *Genetics* 2022;220:iyac001. https://doi.org/10.1093/genetics/iyac001

2. International Society for Biocuration. Biocuration: distilling data into knowledge. *PLoS Biol* 2018;16:e2002846. https://doi.org/10.1371/journal.pbio.2002846

3. Hayden EC. Concern over funding cuts for model organism databases. *Nature* 2016. https://doi.org/10.1038/nature.2016.20134

4. Kaiser J. Funding for key data resources in jeopardy. *Science* 2016;351:14. https://doi.org/10.1126/science.351.6268.14

5. Alliance of Genome Resources Consortium. The Alliance of Genome Resources: Building a Modern Data Ecosystem for Model Organism Databases. *Genetics* 2019;213:1189-1196. https://doi.org/10.1534/genetics.119.302523

6. Naithani S, Gupta P, Preece J et al. Involving community in genes and pathway curation. *Database (Oxford)* 2019;2019:bay146. https://doi.org/10.1093/database/bay146

7. Imker H. Who Bears the Burden of Long-Lived Molecular Biology Databases? *Data Sci J* 2020;19:8. https://doi.org/10.5334/dsj-2020-008

8. Alliance of Genome Resources Consortium. Updates to the Alliance of Genome Resources central infrastructure. *Genetics*2024;227:iyae049. https://doi.org/10.1093/genetics/iyae049

9. Poveda L, Farrell G, Tosatto S et al. The missing link in FAIR data policy: data resources. *Zenodo* 2025. https://doi.org/10.5281/zenodo.15724104

10. Carme P, Rutherford K, Bähler J et al. PomBase in 2026: expanding knowledge, modeling connections. *Genetics*2026;232:iyag001. https://doi.org/10.1093/genetics/iyag001

11. Berardini TZ, Li D, Muller R et al. Assessment of community-submitted ontology annotations from a novel database-journal partnership. *Database (Oxford)* 2012;2012:bas030. https://doi.org/10.1093/database/bas030

12. Bunt SM, Grumbling GB, Field HI et al.; FlyBase Consortium. Directly e-mailing authors of newly published papers encourages community curation. *Database (Oxford)* 2012;2012:bas024. https://doi.org/10.1093/database/bas024

13. Arnaboldi V, Raciti D, Van Auken K et al. Text mining meets community curation: a newly designed curation platform to improve author experience and participation at WormBase. *Database (Oxford)* 2020;2020:baaa006. https://doi.org/10.1093/database/baaa006

14. Lock A, Harris MA, Rutherford K et al. Community curation in PomBase: enabling fission yeast experts to provide detailed, standardized, sharable annotation from research publications. *Database (Oxford)* 2020;2020:baaa028. https://doi.org/10.1093/database/baaa028

15. Martens M, Ammar A, Riutta A et al. WikiPathways: connecting communities. *Nucleic Acids Res* 2021;49:D613-D621. https://doi.org/10.1093/nar/gkaa1024

16. Wang Y, Wang Q, Huang H et al.; UniProt Consortium. A crowdsourcing open platform for literature curation in UniProt. *PLoS Biol* 2021;19:e3001464. https://doi.org/10.1371/journal.pbio.3001464

17. Reiser L, Bakker E, Subramaniam S et al. The Arabidopsis Information Resource in 2024. *Genetics* 2024;227:iyae027. https://doi.org/10.1093/genetics/iyae027

18. Nugnes MV, Bouhraoua KEA, Zoubiri M et al.; DisProt Consortium. DisProt in 2026: enhancing intrinsically disordered proteins accessibility, deposition, and annotation. *Nucleic Acids Res* 2026;54:D383-D392. https://doi.org/10.1093/nar/gkaf1175

19. Sternberg PW, Van Auken K, Wang Q et al. WormBase 2024: status and transitioning to Alliance infrastructure. *Genetics*2024;227:iyae050. https://doi.org/10.1093/genetics/iyae050

20. UniProt Consortium. UniProt: the Universal Protein Knowledgebase in 2025. *Nucleic Acids Res* 2025;53:D609-D617. https://doi.org/10.1093/nar/gkae1010

21. York WS, Mazumder R, Ranzinger R et al. GlyGen: Computational and Informatics Resources for Glycoscience. *Glycobiology*2020;30:72-73. https://doi.org/10.1093/glycob/cwz080

22. Olson A, Kumari S, Wei X et al. Gramene 2025: expanded comparative genomics and pathway resources, integrated search, and pan-genome portals for crop research. *Nucleic Acids Res* 2026;54:D1720-D1732. https://doi.org/10.1093/nar/gkaf1260

23. Gupta P, Elser J, Hooks E et al. Plant Reactome Knowledgebase: empowering plant pathway exploration and OMICS data analysis. *Nucleic Acids Res* 2024;52:D1538-D1547. https://doi.org/10.1093/nar/gkad1052

24. The ClinGen Consortium. The Clinical Genome Resource (ClinGen): Advancing genomic knowledge through global curation. *Genet Med* 2025;27:101228. https://doi.org/10.1016/j.gim.2024.101228

25. Mazandu GK, Hotchkiss J, Nembaware V et al. The Sickle Cell Disease Ontology: recent development and expansion of the universal sickle cell knowledge representation. *Database (Oxford)* 2022;2022:baac014. https://doi.org/10.1093/database/baac014

26. Putman TE, Schaper K, Matentzoglu N et al. The Monarch Initiative in 2024: an analytic platform integrating phenotypes, genes and diseases across species. *Nucleic Acids Res* 2024;52:D938-D949. https://doi.org/10.1093/nar/gkad1082

27. Ragueneau E, Gong C, Sinquin P et al. The Reactome Knowledgebase 2026. *Nucleic Acids Res* 2026;54:D673-D681. https://doi.org/10.1093/nar/gkaf1223

28. Agrawal A, Balcı H, Hanspers K et al. WikiPathways 2024: next generation pathway database. *Nucleic Acids Res* 2024;52:D679-D689. https://doi.org/10.1093/nar/gkad960

29. Ramsey J, McIntosh B, Renfro D et al. Crowdsourcing biocuration: The Community Assessment of Community Annotation with Ontologies (CACAO). *PLoS Comput Biol* 2021;17:e1009463. https://doi.org/10.1371/journal.pcbi.1009463

30. Dahlberg CL, Grove CA, Hulsey-Vincent H et al. Student Annotations of Published Data as a Collaboration between an Online Laboratory Course and the C. elegans Database, WormBase. *J Microbiol Biol Educ* 2021;22:22.1.21. https://doi.org/10.1128/jmbe.v22i1.2331

31. Naithani S, Gupta P, Preece J et al. Plant Reactome: a knowledgebase and resource for comparative pathway analysis. *Nucleic Acids Res* 2020;48:D1093-D1103. https://doi.org/10.1093/nar/gkz996

32. Dunn NA, Unni DR, Diesh C et al. Apollo: Democratizing genome annotation. *PLoS Comput Biol* 2019;15:e1006790. https://doi.org/10.1371/journal.pcbi.1006790

33. Armit C, Tuli MA, Hunter CI. A Decade of GigaScience: GigaDB and the Open Data Movement. *Gigascience* 2022;11:giac053. https://doi.org/10.1093/gigascience/giac053

34. Raciti D, Yook K, Harris TW et al. Micropublication: incentivizing community curation and placing unpublished data into the public domain. *Database (Oxford)* 2018;2018:bay013. https://doi.org/10.1093/database/bay013

35. Rangarajan A, Schedl T, Yook K et al. Toward an interactive article: integrating journals and biological databases. *BMC Bioinformatics* 2011;12:175. https://doi.org/10.1186/1471-2105-12-175

36. Xu W, Gupta A, Jaiswal P et al. Improving publication pipeline with automated biological entity detection and validation service. *Data Inform Manage* 2019;3:3-17.

37. Rosonovski S, Levchenko M, Bhatnagar R et al. Europe PMC in 2023. *Nucleic Acids Res* 2024;52:D1668-D1676. https://doi.org/10.1093/nar/gkad1085

38. Wei CH, Allot A, Lai PT et al. PubTator 3.0: an AI-powered literature resource for unlocking biomedical knowledge. *Nucleic Acids Res* 2024;52:W540-W546. https://doi.org/10.1093/nar/gkae235

39. Allot A, Lee K, Chen Q et al. LitSuggest: a web-based system for literature recommendation and curation using machine learning. *Nucleic Acids Res* 2021;49:W352-W358. https://doi.org/10.1093/nar/gkab326

40. Allot A, Wei CH, Phan L et al. Tracking genetic variants in the biomedical literature using LitVar 2.0. *Nat Genet* 2023;55:901-903. https://doi.org/10.1038/s41588-023-01414-x

41. Dhombres F, Morgan P, Chaudhari BP et al. Prenatal phenotyping: A community effort to enhance the Human Phenotype Ontology. *Am J Med Genet C Semin Med Genet* 2022;190:231-242. https://doi.org/10.1002/ajmg.c.31989

42. Vasilevsky N, Gehrke S, Mullen K et al. Epilepsy disease classification: a community effort to enhance the Mondo Disease Ontology. *Database (Oxford)* 2026;2026:baag004. https://doi.org/10.1093/database/baag004

43. Gene Ontology Consortium. The Gene Ontology knowledgebase in 2026. *Nucleic Acids Res* 2026;54:D1779-D1792. https://doi.org/10.1093/nar/gkaf1292

44. Quaglia F, Chasapi A, Nugnes MV et al. Best practices for the manual curation of intrinsically disordered proteins in DisProt. *Database (Oxford)* 2024;2024:baae009. https://doi.org/10.1093/database/baae009

45. Comeau DC, Wei CH, Islamaj Doğan R et al. PMC text mining subset in BioC: about three million full-text articles and growing. *Bioinformatics* 2019;35:3533-3535. https://doi.org/10.1093/bioinformatics/btz070

46. Rutherford KM, Harris MA, Lock A et al. Canto: an online tool for community literature curation. *Bioinformatics* 2014;30:1791-1792. https://doi.org/10.1093/bioinformatics/btu103

47. Beard N, Bacall F, Nenadic A et al. TeSS: a platform for discovering life-science training opportunities. *Bioinformatics*2020;36:3290-3291. https://doi.org/10.1093/bioinformatics/btaa047

48. Matentzoglu N, Goutte-Gattat D, Tan SZK et al. Ontology Development Kit: a toolkit for building, maintaining and standardizing biomedical ontologies. *Database (Oxford)* 2022;2022:baac087. https://doi.org/10.1093/database/baac087

49. Viteri G, Matthews L, Varusai T et al. Reactome and ORCID-fine-grained credit attribution for community curation. *Database (Oxford)* 2019;2019:baz123. https://doi.org/10.1093/database/baz123

50. Hatos A, Quaglia F, Piovesan D et al. APICURON: a database to credit and acknowledge the work of biocurators. *Database (Oxford)* 2021;2021:baab019. https://doi.org/10.1093/database/baab019

51. Wu G, Matthews L, Boyer N et al. Application of Large Language Models for Annotating Genes into Reactome Pathways. *bioRxiv* 2025:2025.12.20.695723. https://doi.org/10.64898/2025.12.20.695723

52. Caufield JH, Hegde H, Emonet V et al. Structured Prompt Interrogation and Recursive Extraction of Semantics (SPIRES): a method for populating knowledge bases using zero-shot learning. *Bioinformatics* 2024;40:btae104. https://doi.org/10.1093/bioinformatics/btae104

53. Caufield H, Kroll C, O'Neil ST et al. CurateGPT: A flexible language-model assisted biocuration tool. *arXiv* 2024. https://arxiv.org/abs/2411.00046

54. Gargano MA, Matentzoglu N, Coleman B et al. The Human Phenotype Ontology in 2024: phenotypes around the world. *Nucleic Acids Res* 2024;52:D1333-D1346. https://doi.org/10.1093/nar/gkad1005

55. Lai PT, Coudert E, Aimo L et al. EnzChemRED, a rich enzyme chemistry relation extraction dataset. *Sci Data* 2024;11:982. https://doi.org/10.1038/s41597-024-03835-7

56. Bansal P, Morgat A, Axelsen KB et al. Rhea, the reaction knowledgebase in 2022. *Nucleic Acids Res* 2022;50:D693-D700. https://doi.org/10.1093/nar/gkab1016

57. Cerezo M, Sollis E, Ji Y et al. The NHGRI-EBI GWAS Catalog: standards for reusability, sustainability and diversity. *Nucleic Acids Res* 2025;53:D998-D1005. https://doi.org/10.1093/nar/gkae1070

58. Tirunagari S, Saha S, Venkatesan A et al. Lit-OTAR framework for extracting biological evidences from literature. *Bioinformatics* 2025;41:btaf113. https://doi.org/10.1093/bioinformatics/btaf113

59. Poux S, Arighi CN, Magrane M et al.; UniProt Consortium. On expert curation and scalability: UniProtKB/Swiss-Prot as a case study. *Bioinformatics* 2017;33:3454-3460. https://doi.org/10.1093/bioinformatics/btx439

60. Wiegers TC, Davis AP, Wiegers J et al. Integrating AI-powered text mining from PubTator into the manual curation workflow at the Comparative Toxicogenomics Database. *Database (Oxford)* 2025;2025:baaf013. https://doi.org/10.1093/database/baaf013

61. Gabriel L, Becker F, Hoff KJ et al. Tiberius: end-to-end deep learning with an HMM for gene prediction. *Bioinformatics*2024;40:btae685. https://doi.org/10.1093/bioinformatics/btae685

62. Holst F, Bolger AM, Kindel F et al. Helixer: ab initio prediction of primary eukaryotic gene models combining deep learning and a hidden Markov model. *Nat Methods* 2026;23:732-739. https://doi.org/10.1038/s41592-025-02939-1

**Table 1.** Summary of the workshop participating groups, the resources' URL, and the group's representatives.

| Group | URL | Participant(s) |
|---|---|---|
| **Model Organism Knowledgebases** | | |
| **ACKnowledge** | https://wormbase.github.io/ACKnowledge/ | Daniela Raciti, Kimberly Van Auken |
| **PomBase** | https://pombase.org/ | Valerie Wood |
| **TAIR** | https://phoenixbioinfo.org/ | Leonore Reiser |
| **WormBase** | https://wormbase.org/ | Chris Grove |
| **Multi-organism Knowledgebases** | | |
| **DisProt** | https://disprot.org/ | Maria Victoria Nugnes |
| **GlyGen** | https://www.glygen.org/ | Rene Ranzinger |
| **Gramene** | https://gramene.org/ | Sushma Naithani |
| **UniProt** | https://www.uniprot.org/ | Karen Ross |
| **Clinical/Disease Knowledgebases** | | |
| **ClinGen** | https://clinicalgenome.org/ | Courtney Thaxton |
| **Monarch Initiative** | https://monarchinitiative.org | Sabrina Toro |
| **Sickle Cell Disease Ontology** | https://scdontology.h3abionet.org/ | Jade Hotchkiss |
| **Pathway Knowledgebases** | | |
| **Plant Reactome** | https://plantreactome.gramene.org/ | Sushma Naithani |
| **Reactome** | https://reactome.org | Nancy Li |
| **WikiPathways** | https://www.wikipathways.org/ | Susan Coort |
| **Undergraduate Academic Coursework** | | |
| **CACAO** | https://cacao.wiki/ | Jolene Ramsey, Deborah Siegele |
| **Plant Reactome** | https://plantreactome.gramene.org/ | Sushma Naithani |
| **WormBase** | https://wormbase.org/ | Chris Grove |
| **Genome Annotation** | | |
| **Apollo** | https://genomearchitect.readthedocs.io/ | Garrett Stevens |
| **Scientific Publishers** | | |
| **GigaScience** | https://academic.oup.com/gigascience | Bastien Molcrette |
| **microPublication Biology** | https://www.micropublication.org/ | Daniela Raciti, Karen Yook |
| **Literature Databases** | | |
| **Europe PMC** | https://europepmc.org/ | Matt Jeffryes |
| **PubTator** | https://www.ncbi.nlm.nih.gov/research/pubtator3/ | Zhiyong Lu |
| **LitVar** | https://www.ncbi.nlm.nih.gov/research/litvar2/ | |

**Table 2.** Recruitment strategies adopted by the participating resources.

| Recruitment Strategy | Groups |
| --- | --- |
| **Author outreach / Email contact** | ACKnowledge, PomBase, WormBase, Reactome |
| **Self-directed / Website** | ACKnowledge, Clingen, DisProt, Europe PMC, Monarch Initiative, PubTator, PomBase, Sickle Cell Disease Ontology, UniProt, WikiPathways, WormBase, Reactome, Plant Reactome |
| **Workshops / Conferences / Webinars** | Apollo, ClinGen, DisProt, Monarch Initiative, Sickle Cell Disease Ontology, WikiPathways, Reactome, Plant Reactome |
| **Educational credits (undergraduate and continuing education)** | CACAO, ClinGen, Plant Reactome, WormBase |
| **Professional networks / Expert groups** | ClinGen, DisProt, GlyGen, Gramene, Monarch Initiative, Sickle Cell Disease Ontology, Plant Reactome, Planteome, WikiPathways |
| **Journal submission / Publication-integrated curation** | GigaScience, Genetics Society of America journals, microPublication Biology, TAIR, Plant Cell, Plant Physiology |
| **Social media** | ClinGen, Reactome |
| **Workflow integration / Embedded in tools** | Apollo, WikiPathways |

**Table 3.** Types of curated data.

| Type of curated data | Groups |
|---|---|
| **Clinical variant curation** | Clingen |
| **Data publication support** | Genetics Society of America journals, GigaScience, microPublication Biology |
| **Entity recognition via text mining** | ACKnowledge, Genetics Society of America journals, Europe PMC, microPublication Biology, PubTator, LitVar |
| **Gene/Protein Function** | CACAO, DisProt, TAIR, PomBase, UniProt, WormBase, Plant Reactome |
| **Genome annotation and additional enrichment** | Apollo, TAIR, Gramene |
| **Glycoscience** | Glygen |
| **Ontology curation** | Monarch Initiative, Sickle Cell Disease Ontology |
| **Pathway curation** | Reactome, WikiPathways, Plant Reactome |

**Table 4.** Quality control methods employed by the participating groups.

| Quality Control method | | Groups |
|---|---|---|
| **Data validation / checks** (e.g. autocompletes, history tracking) | Automated | ACKnowledge, Apollo, CACAO, Europe PMC, Glygen, Monarch Initiative, PubTator, LitVar, Plant Reactome, PomBase, Reactome, WikiPathways |
| **Professional biocuration/expert review** | Human-based | ACKnowledge, DisProt, microPublication Biology, Monarch Initiative, PomBase, Reactome, TAIR, UniProt, WormBase, Plant Reactome |
| **Peer review / panel verification / consensus** | Human-based | CACAO, ClinGen, Monarch Initiative, Plant Reactome, Reactome, Sickle Cell Disease Ontology, TAIR |
| **QA protocols / community feedback / constrained workflows** | Mixed | DisProt, Monarch Initiative, Plant Reactome, Reactome, TAIR, WikiPathways |
| **Ontology guidelines / rules** | Mixed | Monarch Initiative, Sickle Cell Disease Ontology |

**Table 5.** Incentivization and recognition strategies.

| Incentivation and Recognition | Examples | Groups |
|---|---|---|
| **Acknowledgement on resource website pages** | Contributors listed on homepage, reference pages, curated data pages, person/reviewer pages | DisProt, PomBase, Plant Reactome, Reactome, Sickle Cell Disease Ontology, TAIR, UniProt, WikiPathways, WormBase |
| **Integration with ORCID** | Contributions are cited on contributors ORCID records | APICURON/DisProt, Apollo, GlyGen, Monarch Initiative, Reactome, UniProt, WikiPathways |
| **Prioritization or elevation of community submissions** | Community-curated data is expedited for inclusion in the resource; community curation or review is given a higher rating or status | ACKnowledge, PomBase, Reactome, WormBase |
| **Educational credits and/or training certificates** | Community contributions satisfy continuing or undergraduate education requirements | CACAO, ClinGen, WormBase, Plant Reactome |
| **Authorship on publications and meeting presentations** | Resources include community contributors as authors | ClinGen, DisProt, Monarch Initiative, Plant Reactome, Reactome, Sickle Cell Disease Ontology |
| **Timely and FAIR publications** | Journals work with authors to validate and curate data as part of the publication process | Gigascience, Genetics Society of America journals, microPublication Biology |
| **Rewards or prizes** | Tiered badging system for different levels of contributions; leaderboards; prizes or trophies to top contributors | APICURON/DisProt, ACKnowledge, CACAO, ClinGen, TAIR, WikiPathways, WormBase |

**Table 6.** Automation, machine learning and artificial intelligence applications.

| Application | Methods | Group(s) |
|---|---|---|
| **Identifying literature and topics for curation** | Neural networks | ACKnowledge, ClinGen, DisProt (DisTriage), Reactome, LitSuggest, Rhea, WormBase |
| **Named entity recognition (e.g. gene, variant, strain, etc.)** | String matching (may include secondary rules or filters) | ACKnowledge, ClinGen, Europe PMC, microPublication Biology, PubTator, LitVar, Wikipathways, WormBase |
| **Fact/relation extraction** | Agentic AI | ACKnowledge, PubTator, UniProt, Wikipathways |
| **Ontology term recognition/translation** | String matching | ClinGen, Europe PMC, Monarch Initiative, SCDO |
| **Ontology term creation** | Agentic AI | Monarch Initiative |
| **Genome annotation** | Integration of machine learning-based gene structure predictions | Apollo |

**Table 7.** Sustainability strategies.

| Sustainability Strategy | Group(s) | Examples |
|---|---|---|
| **Diversify and increase funding for professional biocurators and software engineers** | All groups | Embed community curation support in research grants or educational/outreach grants |
| **Integrate community curation into education** | CACAO, ClinGen, WormBase, Plant Reactome | Educational integration ensures ongoing participation and skill development |
| **Partnerships with journals or professional societies for pre-publication curation** | GigaScience, microPublication Biology, Genetics Society of America journals, Plant Cell, Plant Physiology | Supports pre-publication data validation, curation pipelines, and credit for contributors |
| **Centralize and standardize community curation tools, databases, training, and quality control** | All groups | Shared infrastructure reduces redundant development, ensures quality, FAIR principles, and adoption by new groups |
| **Automation / AI / Machine Learning augmentation** | ACKnowledge, Apollo, CACAO, ClinGen, DisProt, Europe PMC, Glygen, NLM (LitSuggest, LitVar, Rhea), microPublication Biology, Monarch Initiative, Plant Reactome, PomBase, PubTator, Reactome, SCDO, Sickle Cell, UniProt, WikiPathways | Supports scalability, quality assurance, and faster literature processing |
| **Collaboration / Cross-group integration** | Monarch Initiative | Promotes shared learning and reduces duplication |